\documentclass[reprint,aps,pra,superscriptaddress,notitlepage]{revtex4-1}
\usepackage{amssymb,amsmath}
\usepackage{graphicx}
\usepackage{dcolumn}
\usepackage{multirow}
\usepackage{color}
\usepackage[english]{babel}
\usepackage[caption=false]{subfig} 

\usepackage{color,soul}  
\sethlcolor{yellow}

\newcommand{\ket}[1]{\ensuremath{|{#1}\rangle}}

\begin{document}

\def\simlt{\mathrel{\lower .3ex \rlap{$\sim$}\raise .5ex \hbox{$<$}}}

\title{\textbf{\fontfamily{phv}\selectfont 
 State-conditional coherent charge qubit oscillations in a Si/SiGe quadruple quantum dot}}
\author{D.\ R.\ Ward}
\thanks{These authors contributed equally to this work.}
\affiliation{Department of Physics, University of Wisconsin-Madison, Madison, WI 53706, USA}
\affiliation{Sandia National Laboratories, Albuquerque, NM 87185, USA}
\author{Dohun Kim}
\thanks{These authors contributed equally to this work.}
\affiliation{Department of Physics, University of Wisconsin-Madison, Madison, WI 53706, USA}
\affiliation{Department of Physics and Astronomy, Seoul National University, Seoul 08826, Republic of Korea}
\author{D.\ E.\ Savage}
\author{M.\ G.\ Lagally}
\affiliation{Department of Materials Science and Engineering, University of Wisconsin-Madison, Madison, WI 53706, USA}
\author{R.\ H.\ Foote}
\author{Mark Friesen}
\author{S.\ N.\ Coppersmith}
\author{M.\ A.\ Eriksson}
\affiliation{Department of Physics, University of Wisconsin-Madison, Madison, WI 53706, USA}

\begin{abstract} 
Universal quantum computation requires high fidelity single qubit rotations and controlled two qubit gates. Along with high fidelity single qubit gates, strong efforts have been made in developing robust two qubit logic gates in electrically-gated quantum dot systems to realize a compact and nano-fabrication-compatible architecture. Here, we perform measurements of state-conditional coherent oscillations of a charge qubit. Using a quadruple quantum dot formed in a Si/SiGe heterostructure, we show the first demonstration of coherent two-axis control of a double quantum dot charge qubit in undoped Si/SiGe, performing Larmor and Ramsey oscillation measurements. We extract the strength of the capacitive coupling between double quantum dots by measuring the detuning energy shift ($\approx$ 75 $\mu$eV) of one double dot depending on the excess charge configuration of the other double dot. We further demonstrate that the strong capacitive coupling allows fast conditional Landau-Zener-Stuckelberg interferences with conditonal $\pi$ phase flip time about 80 ps, showing promising pathways toward multi-qubit entanglement control in semiconductor quantum dots. 
\end{abstract}

\maketitle 

\begin{figure} [t]
\includegraphics[width=0.47\textwidth]{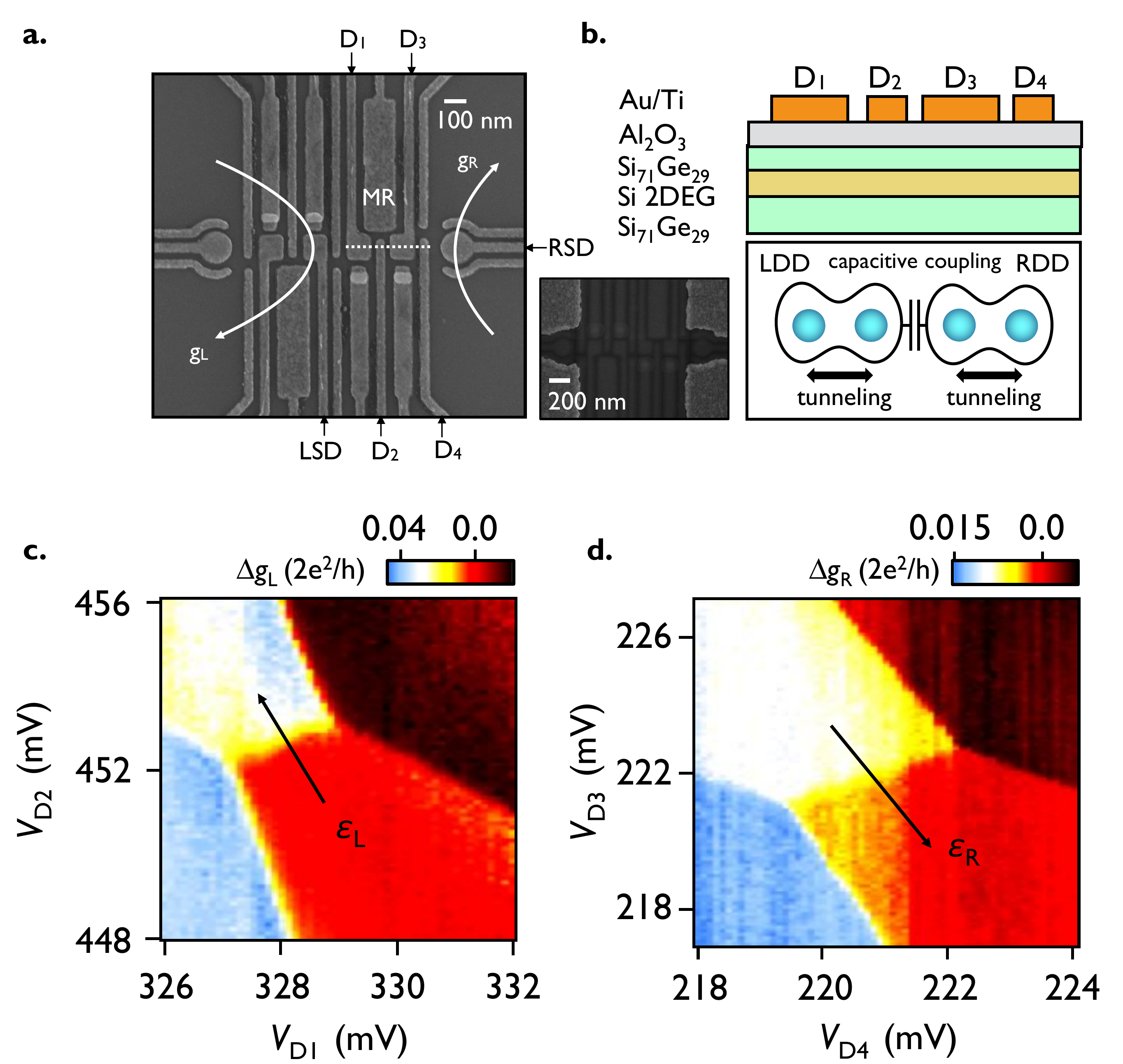}
\caption{\textbf{Si/SiGe device structure and charge stability diagrams of a pair of double quantum dots. a.} Scanning electron microscope image and schematic labeling of a device lithographically identical to the one used in the experiment. For clarity, only the gates in the bottom level are shown in the main panel. The inset to Fig.1a shows the completed device with top level gates. Conductances through the left and right sensor dots (LSD and RSD) were used to monitor the charge occupations in the left and right double dots. \textbf{b.} Schematic cross section through dashed line in Fig. 1a and diagram of a pair of double quantum dots formed under gates from $D_1$ to $D_4$. \textbf{c-d.} Charge stability diagrams of the left (\textbf{c}) and right (\textbf{d}) double dots, measured using conductance of LSD ($g_\text{L}$) and RSD ($g_\text{R}$), respectively. For clarity, a linear background was subtracted from the raw data and resultant conductance variations ($\Delta g_\text{L}$ and $\Delta g_\text{R}$) are plotted.}
\label{fig1} 
\end{figure}

Since being proposed theoretically,~\cite{Loss:1998p120,Kane:1998p133}, much experimental and theoretical progress has been made towards the development of a scalable quantum computing architecture using electrically gated semiconductor quantum dots~\cite{Hanson:2007p1217,Zwanenburg:2013p961,Elzerman:2004p431,Petta:2005p2180,Koppens:2006p766,Foletti:2009p903,Laird:2010p1985,Gaudreau:2011p54,Pla:2012p489,Medford:2013p654,Buch:2013p2017,Eng:2015p41,Reed:2015unpublished,Martins:2015unpublished,Takeda:2016unpublished}.
Developments in nanofabrication and high fidelity measurement techniques have enabled substantial  progress towards realizing two qubit gates in semiconductor quantum dot systems, including the measurement of interqubit capacitive coupling strength~\cite{Li:2015p7681}, conditional coherent exchange oscillations in GaAs singlet-triplet qubits \cite{Shinkai:2009p056802,Petersson:2009p016805,VanWeperen:2011p030506}, and demonstration of nontrivial entanglement of capacitively coupled GaAs singlet-triplet qubits~\cite{Shulman:2012p202}. More recently, high fidelity conditional gate operation has been demonstrated in silicon-based quantum dot spin qubits \cite{Veldhorst:2014p981,Veldhorst:2015p410}, harnessing the substantial improvement in coherence time achieved by isotopic purification of $^{28}\text{Si}$.  

Improving gate speed provides an alternative route to realize high fidelity single and multi qubit gates. Intensive efforts to realize fast manipulation of semiconductor spin qubits by mixing the spin degrees of freedom with charge degrees of freedom have been made \cite{Petta:2005p2180,Wu:2014p11938,Berg:2013p066806,Medford:2013p654}.  Moreover, solely relying on the electric field control, both non-adiabatic coherent control and resonant microwave-driven gate operations have been demonstrated both in GaAs \cite{Petersson:2010p246804,Cao:2013p1401,Cao:2016p086801} and natural Si \cite{Shi:2013p075416, Kim:2015p243} charge and spin qubits, with typical coherence times of the order of 100 ps to 10 ns. Measured interdot capacitive couplings exceed 20 GHz \cite{Li:2015p7681,Shinkai:2009p056802,Petersson:2009p016805,VanWeperen:2011p030506}, so fast multi qubit-gate operations using charge qubits can be expected. 

\begin{figure*}
\includegraphics[width=1\textwidth]{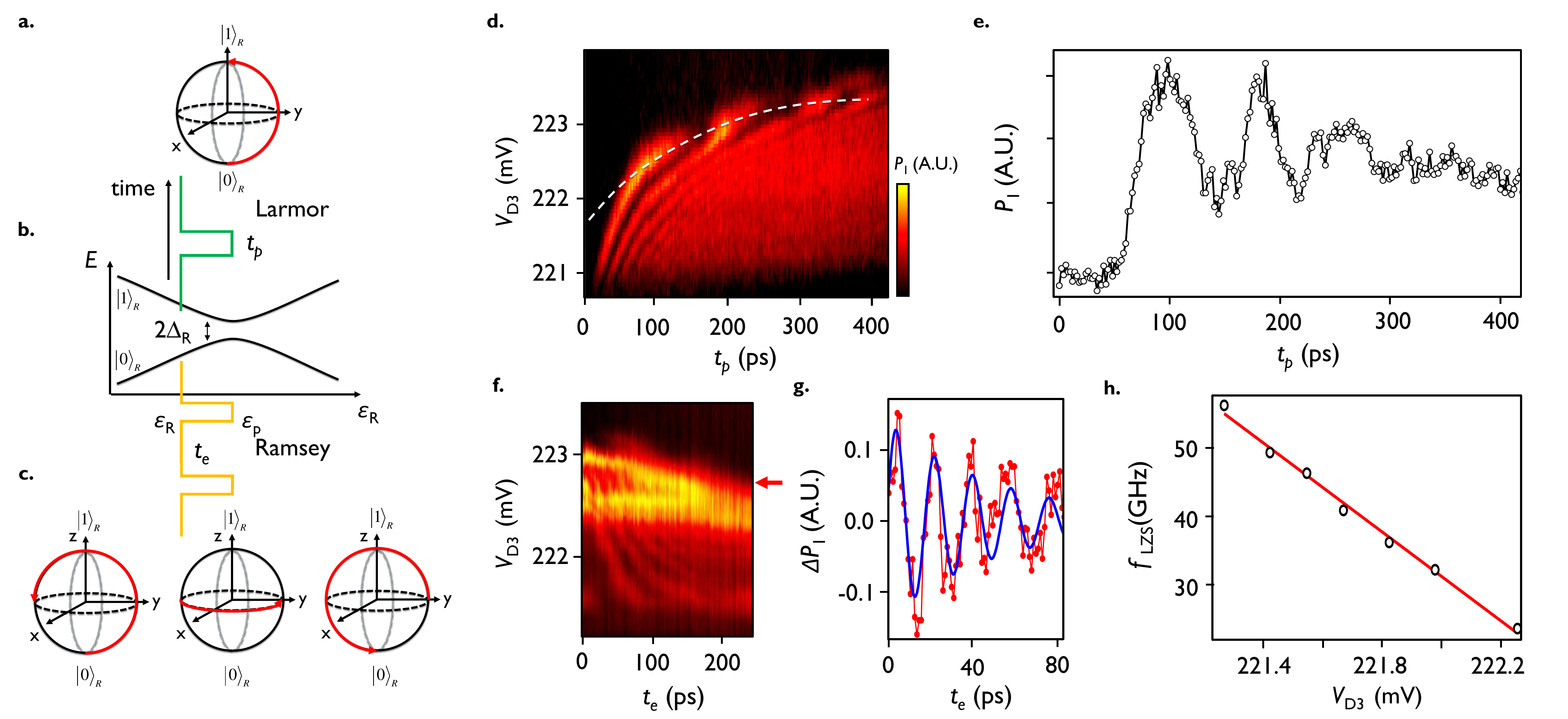}
\caption{\textbf{Demonstration of two-axis control of an undoped Si/SiGe charge qubit formed in the right double dot. a.} Time evolution of the Bloch vector during a non-adiabatic DC-pulsed gate (Larmor oscillation). An abrupt change in the detuning $\varepsilon_\text{R}$ from a negative value, where state $\ket{0}_\text{R}$ is the eigenstate of the Hamiltonian, to $\varepsilon_\text{P}$=0 induces a rotation of the state around the X-axis of the Bloch sphere. \textbf{b.} Schematic energy level diagram of a double dot charge qubit with the pulse sequences for Larmor (X-axis rotation, green) and Ramsey (Z-axis rotation, orange) oscillation measurements.  \textbf{c.} Schematic time evolution of Bloch vector during a Ramsey fringe measurement pulse sequence. A $X_{3\pi/2}$ pulse is applied to initialize the state on the XY plane of the Bloch sphere, and the state then evolves freely around the Z-axis for evolution time $t_\text{e}$ with the rate $E_\text{01,R}/h$ determined by the right qubit energy spliting $E_\text{01,R}$ = $\sqrt{{\varepsilon_\text{R}}^2+({2\Delta_\text{R}})^2}$. A second $X_{3\pi/2}$ pulse maps the Y-axis to the Z-axis, and the average charge occupation is measured via the conductance change of the RSD (see Fig.1a) \textbf{d.} Coherent oscillation of uncalibrated probability $P_{1}$ in arbitrary units corresponding to Larmor oscillations (X-axis rotations) as a function of voltage on gate $D_{3}$, $V_\text{D3}$ and pulse duration $t_{p}$ of a single step pulse (see Fig.2a - green pulse). \textbf{e.} Line cut along the contour shown as the white dashed line in {\bf d.}, which corresponds to $\varepsilon_\text{R}$ = 0, showing coherent Larmor oscillations with coherence time about $T_2^*\approx$ 150 ps. 
\textbf{f.} Demonstration of Z-axis control performed with a Ramsey fringe experiment (orange pulse in Fig.2b). Uncalibrated $P_{1}$ as a function of $V_\text{D3}$ and $t_{e}$. \textbf{g.} Line cut of the Ramsey fringe as a function of $t_\text{e}$. For clarity, a background probability variation of a third order polynomial in time was removed from the raw $P_{1}$, as shown in the Supplementary Fig. S3. The oscillations arise because of rotations of the Bloch vector about the Z-axis of the Bloch sphere. \textbf{h.} Landau-Zener-Stueckelburg (LZS) oscillation frequency {\it f}$_\text{LZS}$ as a function of $V_\text{D3}$ in the regime where the pulse tip detuning $\varepsilon_\text{p} > \Delta_\text{R}$. The red solid line shows a linear fit to  $\Delta f_\text{LZS}\approx \alpha_{\varepsilon_\text{R,D3}} \Delta V_\text{D3}$ with best fit parameter of gate $D_{3}$ lever arm $\alpha_{\varepsilon_\text{R,D3}}\approx$ 32.5 GHz/mV $\approx$ 135 $\mu$eV/mV.}
\label{fig2} 
\end{figure*}

Here, we show fast and charge state conditional coherent manipulation of two strongly coupled double quantum dots. Non-adiabatic pulsed gate techniques allow fast two-axis control of a double dot charge qubit formed in an undoped Si/SiGe heterostructure in the accumulation mode. Furthermore, we show that strong capacitive coupling ($>$ 18 GHz) between two sets of double quantum dots enables charge state conditional coherent Landau-Zener-Stuckelberg interference with a conditional $\pi$ phase flip time about 80 ps, showing promising progress toward realizing high-fidelity two qubit control.

\noindent\textbf{Results} 

We study a linear quadruple quantum dot formed in an undoped Si/SiGe heterostructure, as shown in Fig. 1a. The dots are formed under the gates $D_{1}$ through $D_{4}$, as shown in Fig. 1b, and for the experiments we report here, it is useful to describe the quadruple quantum dot as a pair of double quantum dots. The right double dot (RDD), formed under the gates $D_{3}$ and $D_{4}$, forms a charge qubit that will be manipulated coherently based on the charge state of the left double dot (LDD), which is formed under gates $D_{1}$ and $D_{2}$. Charge sensing is performed by two charge sensing quantum dots adjacent to the left (LSD) and right (RSD) hand sides of the quadruple dot array. The location of sensor RSD is close to the position that would naively be expected by examination of Fig.~1a; to improve its charge sensitivity, sensor LSD is shifted to a position very close to the quadruple dot by careful tuning of the large number of gate voltages available on that side of the device. We monitor changes in the conductances $g_\text{L}$ and $g_\text{R}$ of sensor dots LSD and RSD, respectively, to monitor the electron occupations of double dots LDD and RDD. Figures 1c and 1d show charge stability diagrams for LDD (\textbf{c}) and RDD (\textbf{d}), demonstrating control of the four dot occupations as a function of the four gate voltages $V_\textrm{D1}$, $V_\textrm{D2}$, $V_\textrm{D3}$, and $V_\textrm{D4}$. As we show in Supplementary Fig. S1c, the tunnel coupling and the capacitive coupling between the LDD and RDD both become negligible when the LDD is in the few electron regime. Thus, we perform here coherent manipulation in the regime for which the LDD has a total electron occupation larger than (10,10).
 
We first show coherent two-axis control of an undoped Si/SiGe double charge qubit formed in the RDD. For this demonstration, the LDD energy detuning $\varepsilon_\text{L}$ is kept $>$300 $\mu$eV so that the LDD charge occupation is not affected by the RDD manipulation pulses. The charge qubit states are defined as $\ket{0}_\text{R}=\ket{L}$ (excess charge is on the left dot) and $\ket{1}_\text{R}=\ket{R}$ (excess charge is on the right dot). The initial qubit state $\ket{0}_\text{R}$ is prepared at negative RDD energy detuning $\varepsilon_\text{R}$. As shown schematically in Fig.~2a-c, non-adiabatic control of the charge qubit is performed using abrupt changes in detuning energy with precise control of the pulse duration time as well as the amplitude. The pulses, generated using a Tektronix AWG70002A arbitrary waveform generator (AWG) with a rise time of 40 ps, are applied to gate $D_{3}$ through a commercial bias tee (Picosecond PulseLabs 5542-219). X-rotations on the Bloch sphere, shown in Fig.~2a, correspond to oscillations between the qubit states $\ket{0}_\text{R}$ and $\ket{1}_\text{R}$. 
Changing the detuning abruptly to $\varepsilon_\text{P} =0 $ yields an 
initial state $\ket{0}_\text{R}$ that is a
superposition of the eigenstates of the Pauli matrix $\sigma_\text{x}$. 
At $\varepsilon=0$ the Hamiltonian is $H=\Delta_\text{R}\sigma_\text{x}$, where $\Delta_\text{R}$ is the tunnel coupling between $D_3$ and $D_4$, so the state evolves periodically in time at the Larmor frequency 2$\Delta_\text{R}/h$, where $h$ is Planck's constant. After a time evolution of duration $t_\text{p}$, the final state is measured by abruptly changing the detuning back to negative $\varepsilon_\text{R}$. We use the difference of conductance of the RSD between $\ket{0}_\text{R}$ and $\ket{1}_\text{R}$ to determine a time averaged signal proportional to the probability $P_\text{1}$ of the state being in $\ket{1}_\text{R}$~\cite{Kim:2014nature}.

Figs.~2d and 2e show coherent oscillations of $P_\text{1}$ resulting from the non-adiabatic pulse sequences described above. In Fig.~2d we plot $P_\text{1}$ as a function of $t_\text{p}$ and the gate voltage $V_\text{D3}$, the latter of which determines the base level of $\varepsilon_\text{R}$. In order to overcome a sampling time limitation of our AWG, we modified the pulse generation scheme to allow sub-picosecond timing resolution (see Supplementary Fig.~S2). In Fig.~2d, the path of the pulse tip detuning $\varepsilon_\text{P}=0$ is curved (white dashed curve in Fig.~2d), most likely due to the finite rise time of the pulse and frequency-dependent attenuation in the microwave coaxial cable~\cite{Petersson:2010p246804}. Fig.~2e shows a line cut through the path corresponding to $\varepsilon_\text{P}=0$, revealing periodic oscillations in $P_\text{1}$ at a frequency of order 10~GHz, corresponding to $\Delta_\text{R}/h\simeq 5$~GHz. We typically observe beating of the oscillations after $t_\text{p}$ = 300 ps. This likely arises because of the superposition of a reflected part of the pulse with the original pulse, modifying the detuning amplitude \cite{Petersson:2010p246804, Kim:2014nature}.

The high frequency oscillations of $P_\text{1}$ in Fig. 2d for $V_\text{D3} <$ 222 meV arise from coherent Landau-Zener-Stueckelberg (LZS) interference patterns~\cite{Shevchenko:2010p1, Stehlik:2012p121303}.  As $V_\text{D3}$ becomes less positive in Fig.~2d, the pulse tip detuning enters the regime $\varepsilon_\text{P} > 0$, where the interdot tunnel coupling acts as a beam splitter \cite{Petta:2010p669, Shi:2014p3020}. Here, the splitting ratio between the upper and lower branches of the charge qubit dispersion is determined by the detuning ramp rate in comparison with the tunnel coupling.  On the return edge of the pulse, the two different trajectories returning through the beamsplitter at $\varepsilon_\text{P}=0$ can coherently interfere.

The measurement of qubit state rotations about the Z-axis on the Bloch sphere, shown schematically in Fig.~2c, can be performed using two $X_{3\pi/2}$ pulses. The qubit state is first prepared in the state $\ket{-Y}_\text{R}=\sqrt{1/2}(\ket{0}_\text{R}-i\ket{1}_\text{R})$, by initializing to state $\ket{0}_\text{R}$ and by performing an $X_{3\pi/2}$ rotation. The qubit state then acquires a relative phase $\varphi=e^{-i t_\text{e}\Delta E_\text{01,R}/h}$, where $t_\text{e}$ is the time spent between the two $X$ rotations at the base value of the detuning and the qubit energy splitting $E_\text{01,R}$ = $\sqrt{{\varepsilon_\text{R}}^2+({2\Delta_\text{R}})^2}$. This phase evolution corresponds to a rotation of the qubit state around the Z-axis of the Bloch sphere. Figs.~2f and 2g show the resulting quantum oscillations of the qubit state around the Z-axis of the Bloch sphere. In Fig.~2g, the line cut is taken through $V_\text{D3}\sim 222.7$ mV, corresponding to $\varepsilon_\text{P}=0$, and a smooth third order polynomial background oscillation was removed from the raw data for clarity \cite{Dovzhenko:2011p161802,Shi:2013p075416} (see also Supplementary Fig. S3). By fitting the data to an exponentially damped sinusoidal oscillations, we extract the Ramsey fringe oscillation frequency $f_\text{Ramsey}\approx 56$ GHz and a coherence time $T_2^*\sim 51$ ps. The gate voltage dependence of both the LZS interference and the Ramsey fringe frequencies provide accurate measures of the detuning lever arm. Fig.~2h shows the LZS oscillation frequency $f_\text{LZS}$ as a function of $V_\text{D3}$. As these LZS oscillations are measured in the limit $\varepsilon_\text{P} > \Delta_{R}$, we use approximate form of the charge qubit energy level, $E_\text{01,R}=\sqrt{\varepsilon_\text{R}^2+{(2\Delta_\text{R})}^2} \approx \varepsilon_\text{R}=hf_\text{LZS}$, and fit the data to the form $\Delta f_\text{LZS} = \alpha_{\varepsilon_\text{R,D3}} \Delta V_\text{D3}$ to determine the gate lever arm $\alpha_{\varepsilon_\text{R,D3}}\approx$32.5 GHz/mV$\approx$135$~\mu$eV/mV.

\begin{figure}[t]
\includegraphics[width=0.47\textwidth]{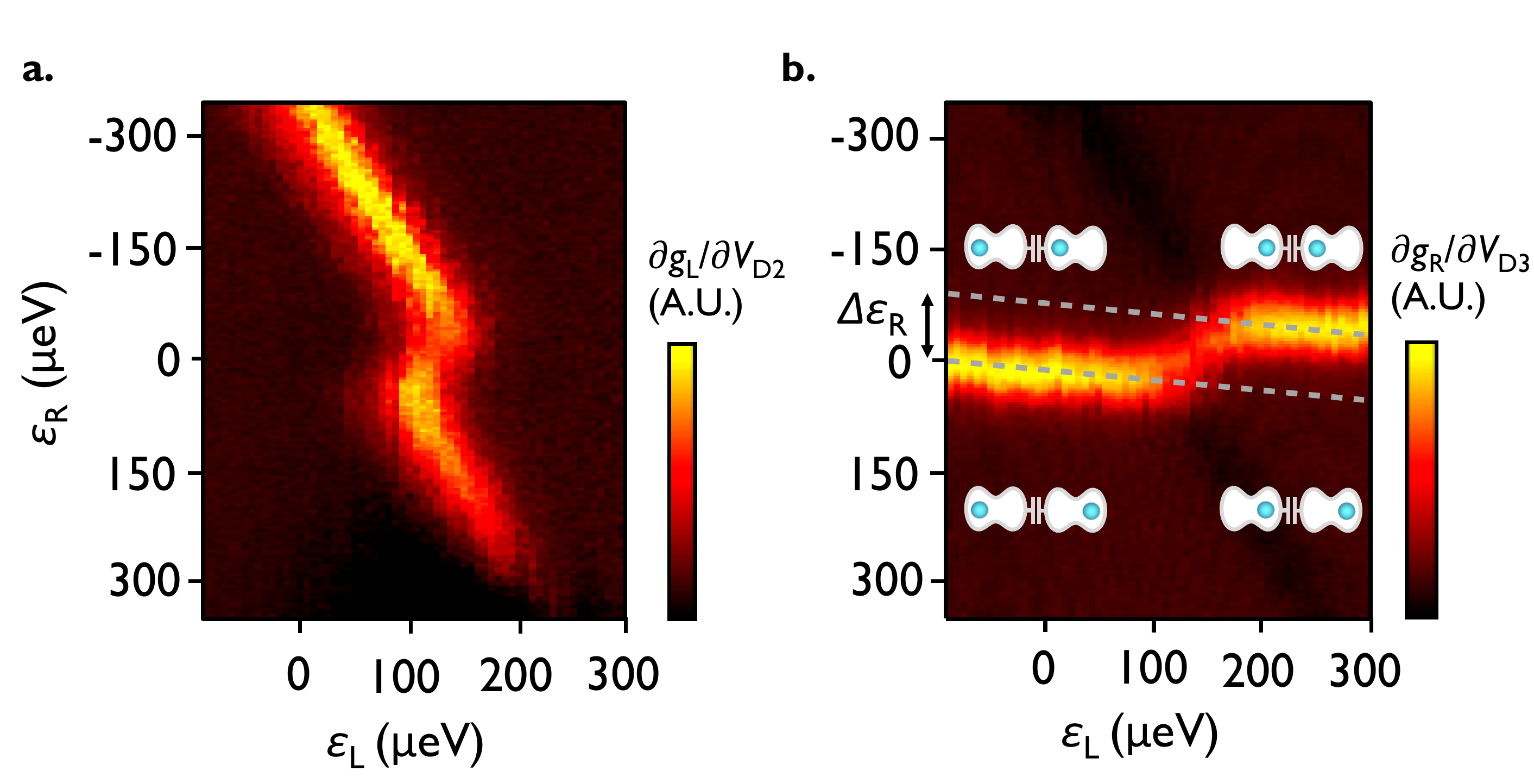}
\caption{\textbf{Measurement of capacitive coupling between two sets of double quantum dots. a-b.} The left and right double dot polarization lines are characterized by measuring the LSD (\textbf{a}) and RSD (\textbf{b}) differential conductances as functions of the detunings $\varepsilon_\text{L}$ and $\varepsilon_\text{R}$. A polarization line is identified by its large differential conductance. In \textbf{b}, a schematic diagram indicates the location of the excess charge in the capacitively coupled double quantum dots. The gray dashed lines represent the shift of RDD polarization line detuning ($\Delta \varepsilon_\text{R}\approx$ 75 $\mu$eV) due to one electron moving from the left dot to the right dot in the LDD. }
\label{fig3} 
\end{figure}

\begin{figure*}
\includegraphics[width=1\textwidth]{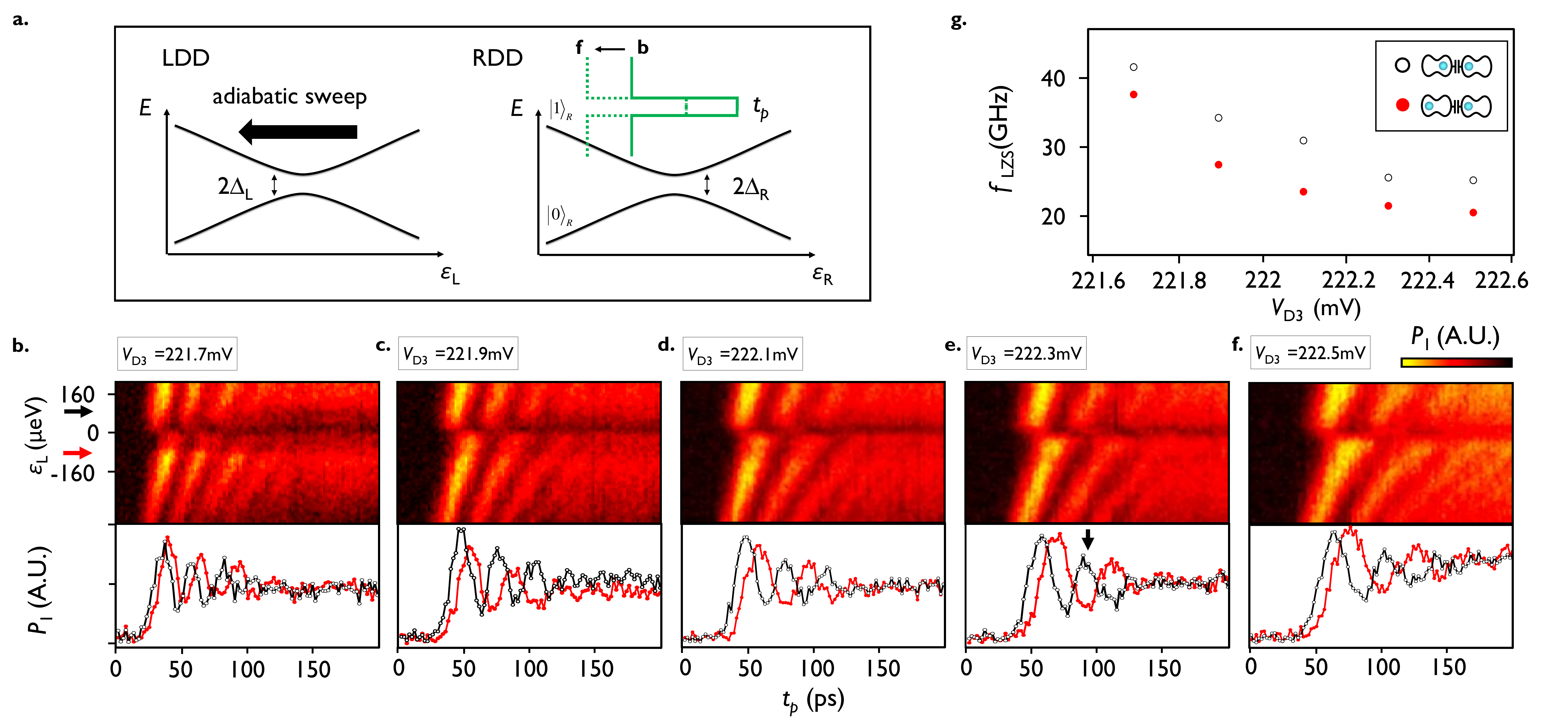}
\caption{\textbf{Charge state conditional coherent quantum interference. a.} Schematic diagram of the pulse sequences used for the measurement of Landau-Zener-Stuckelberg (LZS) quantum interference in the right double dot (RDD) as a function of the detuning of the left double dot (LDD). \textbf{b-f.} Coherent LZS oscillations of uncalibrated $P_{1}$ as a function of the LDD detuning $\varepsilon_\text{L}$ and pulse duration $t_{p}$ with fixed base level of $V_\text{D3}$ = 221.7 (\textbf{b}), 221.9 (\textbf{c}), 222.1 (\textbf{d}), 222.3 (\textbf{e}), and 222.5~mV (\textbf{f}). 
\textbf{g.} LZS interference frequencies, $f_\text{LZS}$, with   $(0,1)_\text{L}$ - (black) and $(1,0)_\text{L}$ - (red) excess charge ground state of the LDD as functions of $V_\text{D3}$.  
The black arrow in {\bf e} highlights the shift in the location of the peak in the probability that corresponds to a conditional $\pi$ phase rotation achieved in $\sim$80~ps.
}
\label{fig4} 
\end{figure*}

We now 
discuss the measurement of the capacitive coupling between the double quantum dots. With the detuning lever arm calibrated as described above, along with mutual capacitances between the gates, the coupling strength can be measured by sweeping $\varepsilon_\text{L}$ and $\varepsilon_\text{R}$ through the LDD and RDD charge degeneracy points. Fig.~3 shows the LDD and RDD polarization lines, characterized by measuring the differential conductance of the left and right sensors, LSD (Fig.~3a) and RSD (Fig.~3b), as
functions of 
the two critical variables, the detuning parameters for the LDD and RDD: $\varepsilon_\text{L}$ and $\varepsilon_\text{R}$.
We sweep $\varepsilon_\text{L}$ and $\varepsilon_\text{R}$ by controlling the voltages on ($V_\text{D1}$, $V_\text{D2}$) and ($V_\text{D3}$, $V_\text{D4}$), respectively.
The positions of the excess charges (the electrons in each double dot that are free to move) are shown schematically in the inset to Fig.~3b.  The coupled charge stability diagram reveals the four possible ground state charge configurations for an extra electron in each of the two double dots.
The gray dashed lines in Fig.~3b show the RDD detuning energy shift ($\Delta\varepsilon_\text{R}$) arising from the movement of a single electron from left to right in the LDD. The shift in this line is a direct measure of the energy shift in the RDD resulting from the capacitive coupling between the two double dots. From the energy calibrations reported above, we extract $\Delta\varepsilon_\text{R}\approx$ 75 $\mu$eV $\approx$ 18.3 GHz.  This energy shift is the available detuning modulation for the performance of two-qubit gates in quantum dots of a size and separation similar to those studied here.

We now show that the capacitive coupling demonstrated above enables fast charge-state-conditional phase evolution of a quantum dot charge qubit. We study LZS oscillations in the RDD in the presence of a perturbation from the excess charge in the LDD. Fig.~4a illustrates schematically the pulse sequence used for this experiment.  The base detuning for the short pulse, shown in green in Fig.~4a, is controlled using $V_\text{D3}$, and in Figs. 4b to 4f, we vary $V_\text{D3}$ from $221.7$ to $222.5$ in steps of 0.2 mV.  The effect of these steps is to change the energy in detuning of the tip of the fast pulse, thus changing the frequency of the LZS oscillations. For each of these LZS oscillation measurements, $\varepsilon_\text{L}$ is swept from +180 to -320 $\mu$eV (the vertical axis in Figs. 4b-f), in order to change the excess charge occupation of the left double dot from $(0,1)_\text{L}$ to $(1,0)_\text{L}$ --- this change occurs quite abruptly at zero detuning of the LDD, as can be observed in Figs.~4b-f. The interference pattern shows two characteristic features as a function of $\varepsilon_\text{L}$.  First, there is a continuous and gradual increase in frequency as a function of $\varepsilon_\text{L}$, arising from the capacitive coupling between the gates above LDD and the RDD. In addition, there is a sudden decrease in frequency in Figs. 4b to 4f as $\varepsilon_\text{L}$ takes the LDD from positive to negative detuning. This decrease in frequency reflects the decreased $\varepsilon_\text{P}$ that the LZS pulse tip reaches, because of the effective change in baseline detuning energy arising from the one electron charge transition in the LDD. The bottom panels of Figs. 4b to 4f show line cuts of the LZS oscillations in the RDD for the $(0,1)_\text{L}$ LDD ground state (black) and $(1,0)_\text{L}$ LDD ground state (red, see black and red arrows in Fig. 4b). Clearly a phase change arises from the motion of one electron in the LDD, and we achieve a conditional $\pi$ phase flip in a time $t_\text{p}$ as short as 80 ps, as indicated by the black arrow in Fig.~4e. Fig.~4g shows the difference in $f_\text{LZS}$ between the cases when the LDD ground state is $(0,1)$ (black circles) and when this ground state is $(1,0)$ (red circles).  This difference in frequency ranges from 7 to 10 GHz and can be used to infer the speed of a conditional phase (CPHASE) gate if full control over qubits in both the left and right double dots is realized in the future. We emphasize that the frequency changes observed here arise from competing effects; $f_\text{LZS}$ of the RDD increases as we change a gate voltage to increase $\varepsilon_\text{L}$, whereas $f_\text{LZS}$ decreases as we cross zero detuning in the LDD, resulting in the motion of a single electron charge. Since we take line cuts at $\varepsilon_\text{L}\approx\pm$80~$\mu$eV to clearly show LZS oscillations in the $(0,1)_\text{L}$ and $(1,0)_\text{L}$ ground states, we believe that using LDD detuning pulse amplitude $<$~160$\mu$eV, when LDD coherent manipulation is realized, can lead to a faster conditional phase gate than estimated here.

\noindent \textbf{Discussion}

Using strong capacitive coupling between two double quantum dots in a linear quad dot array geometry ($\approx$18 GHz), we achieve fast charge state conditional coherent oscillations with a conditional phase flip time of 80 ps, demonstrating the key physical interaction necessary for a two-qubit CPHASE gate. Moreover, because we measure single qubit X (Larmor) and Z (Ramsey) rotations with rotation frequencies also on the order of 10 GHz, one can envision fast universal quantum logic gates in semiconductor charge qubits. Resonant microwave control is also plausible \cite{Kim:2015p243}, in which case a two-qubit controlled not gate (CNOT) can be implemented \cite{Li:2015p7681,Veldhorst:2015p410}. We stress however that the full demonstration of two qubit gates remains as a challenge, as in this work coherent control of the LDD could not be achieved. A more compact gate geometry, for example, using an overlapped Al/$\text{Al}_{2}\text{O}_{3}$ gate structure \cite{Veldhorst:2014p981, Zajac:2015p223507} can be considered, in order to allow tunability as well as strong confinement and large tunnel coupling strength down to the single electron regime.   
 

\noindent \textbf{Methods}

{\it Fabrication:}  The device heterostructure was grown using chemical vapor deposition (CVD) on commercially available SiGe substrates with a 29\% Ge composition.
The CVD growth sequence from the starting substrate was deposition of a strain-matched SiGe buffer layer followed by deposition of a 12~nm thick strained Si well.  The well was capped by deposition of a 50~nm of SiGe, followed by a few nanometers of sacrificial strained Si to cap the heterostructure.

Devices were fabricated using a combination of electron beam lithography (EBL) and photolithography.  The device nanostructure was fabricated in two layers starting on a 15~nm gate dielectric of $\text{Al}_{2}\text{O}_{3}$ deposited by atomic layer deposition (ALD).  The first layer of control gates was patterned in two EBL/metallization steps to improve the gate density and metallized with Ti/Au.  The second reservoir gate layer (see inset to Fig. 1a) is isolated from the first with another 80~nm layer of $\text{Al}_{2}\text{O}_{3}$ grown via ALD.  The second gate layer was also metallized with Ti/Au.  A third layer of $\text{Al}_{2}\text{O}_{3}$ was deposited over the second gate layer to protect the gates during subsequent fabrication steps.  Ohmic contacts were fabricated using annealed P+ ion implants.

{\it Measurement:} The charge stability diagrams of the LDD and RDD are characterized by measuring the conductance changes through the left and right sensor dots (LSD and RSD respectively, see Fig. 1a), which are operated at a fixed voltage bias of 50~$\mu$V, and the currents are measured with two current preamplifiers (DL Instruments model 1211). Supplementary Fig. S1 provides large scale charge stability diagrams and the positions of charge transitions of the LDD and RDD in the few electron regime used in the present experiment. For the manipulation of the RDD charge qubit, fast voltage pulses with repetition rate of 25~MHz are generated using two channel outputs of a Tektronix~AWG70002A arbitrary waveform generator and are added to the dot-defining dc voltage through a bias tee (Picosecond Pulselabs~5546-107) before being applied to gate $D_{3}$. The conductance change through the right sensor dot (RSD) with and without the manipulation pulses, measured with a lock-in amplifier (EG\&G model~7265), is used to determine the average charge occupation and is converted to the probabilities. For the measurement of changes in charge occupation probabilities resulting from fast manipulation pulses, we modulated the manipulation pulses with a low frequency ($\approx$777~Hz) square wave envelope, similar to the technique we used in previous studies~\cite{Kim:2015p243, Kim:2015p15004}. We compare the measured signal level with the corresponding $\ket{0}_\text{R}$ to $\ket{1}_\text{R}$ charge transition signal level, calibrated by sweeping gate $D_{3}$ and applying a 777~Hz square pulse to gate D$_{3}$ with an amplitude the same as the manipulation pulses. 

\noindent \textbf{Acknowledgements}

This work was supported in part by ARO (W911NF-12-0607), NSF (DMR-1206915, PHY-1104660),
ONR (N00014-15-1-0029), and the Department of Defense. The views and conclusions contained in this document are those of the authors and should not be interpreted as representing the official policies, either expressly or implied, of the US Government.  Work also supported by the Laboratory Directed Research and Development program at Sandia National Laboratories. Sandia National Laboratories is a multi-program laboratory managed and operated by Sandia Corporation, a wholly owned subsidiary of Lockheed Martin Corporation, for the U.S. Department of Energy's National Nuclear Security Administration under contract DE-AC04-94AL85000. Development and maintenance of the growth facilities used for fabricating samples is supported by DOE (DE-FG02-03ER46028). This research utilized NSF-supported shared facilities at the University of Wisconsin-Madison.

\noindent \textbf{Author Contributions}

DRW fabricated the quantum dot device and developed hardware and software for the measurements. DK performed electrical measurements with RHF and analyzed the data with MAE, MF, and SNC. DES and MGL prepared the Si/SiGe heterostructure. All authors contributed to the preparation of the manuscript. 

\noindent \textbf{Additional Information}

Supplementary information accompanies this paper. Correspondence and requests for materials should be addressed to Mark A. Eriksson (maeriksson\emph{@}wisc.edu)

\renewcommand{\theequation}{S\arabic{equation}}
\setcounter{equation}{0}
\renewcommand{\thefigure}{S\arabic{figure}}
\renewcommand{\figurename}{Supplementary Fig.}

\setcounter{figure}{0}
\renewcommand{\thesection}{SUPPLEMENTARY NOTE \arabic{section}}
\setcounter{section}{0}

\section*{Supplementary Information}
\section{Charge stability measurements}
\label{sup:stability}

\begin{figure}[b]
\includegraphics[width=0.47\textwidth]{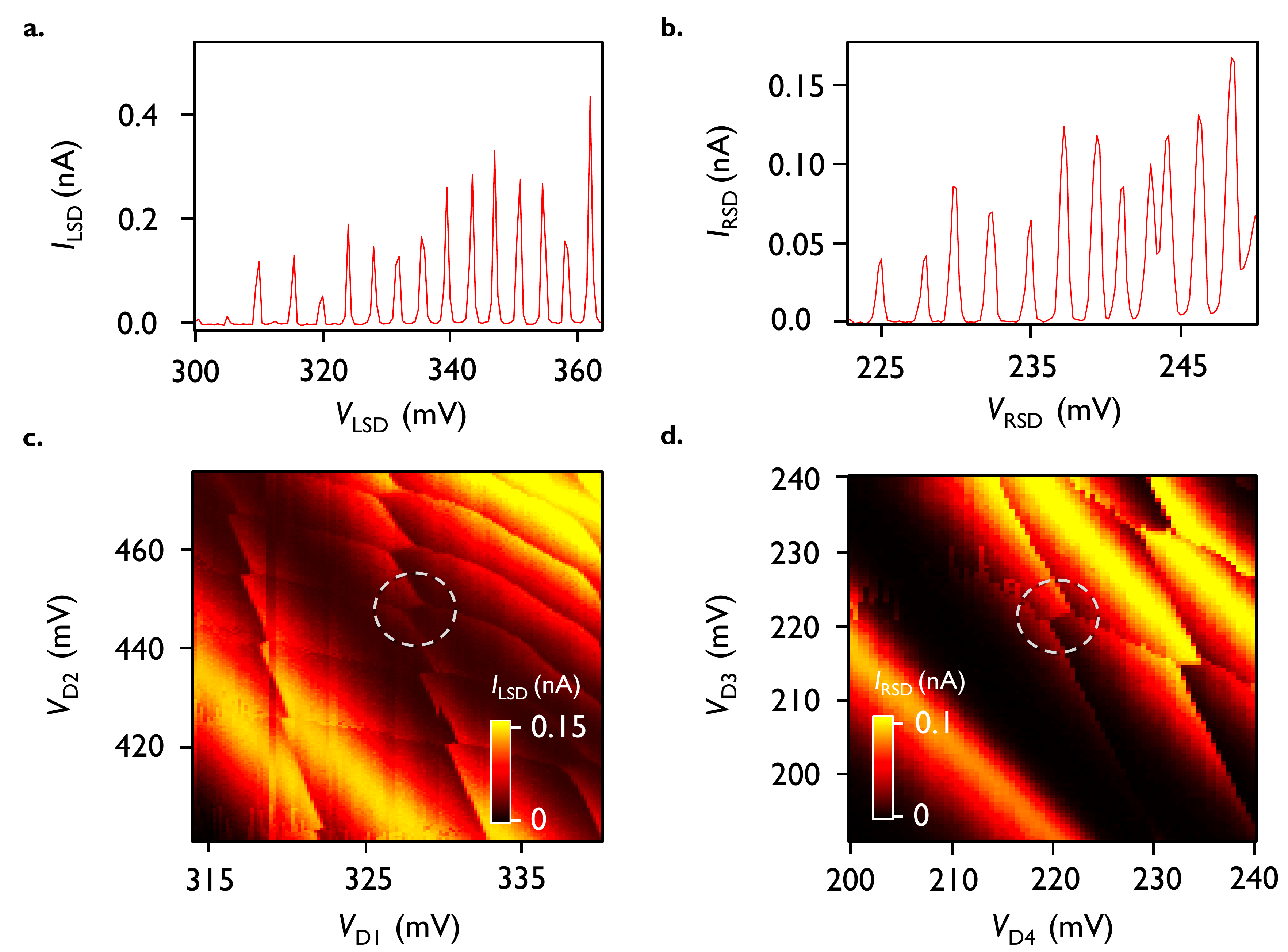}
\caption{ \textbf{Measurement of charge stability diagrams of two pairs of double quantum dots. a-b}, Coulomb peaks obtained by measuring $I_\text{LSD}$ and $I_\text{RSD}$, the current through the single electron transistors formed by the left sensor dot (LSD) (\textbf{a}) and right sensor dot (RSD) (\textbf{b}) as a function of the appropriate gate voltage. \textbf{c-d}, Large scale stability diagrams of left double dot (LDD) (\textbf{c}) and right double dot (RDD) (\textbf{d}) measured by recording the conductance change of LSD and RSD, respectively, as a function of the relevant gate voltages. The white dashed circles show the regions of charge configurations used for the current experiments. }
\label{fig:S1} 
\end{figure}

For measuring changes in charge occupation and charge qubit probabilities, we use two single electron transistors formed on the left and right hand side of the linear quadruple dot array (see Fig.~1a - LSD and RSD in the main text). Supplementary Figs.~1a and 1b show Coulomb blockade peaks  of the charge sensors. By adjusting $V_\text{LSD}$ and $V_\text{RSD}$ to be near maxima of the current through LSD and RSD, the charge occupation in the $D_1$ and $D_2$ ($D_3$ and $D_4$) double dot can be measured using $I_\text{LSD}$ ($I_\text{RSD}$). Supplementary Figs.1c and 1d show the large scale charge stability diagrams of the left double dot (LDD, 1c) and right double dot (RDD,1d), respectively. We find that both the tunnel coupling strength and the inter-double dot capacitive coupling becomes negligible as electrons are removed from the LDD as it can be inferred from the trend of the polarization lines in Supplementary Fig.1c, whereas the tunnel coupling of the RDD can be maintained $>$ 10 GHz down to few electron regime. Thus we choose the LDD electron occupation number to be in the regime $>$ (10, 10). The regions of the charge stability diagrams used for the pulse experiments are denoted as white dashed circles in the Supplementary Figs.~1c and 1d. 
     
\begin{figure}[t]
\includegraphics[width=0.47\textwidth]{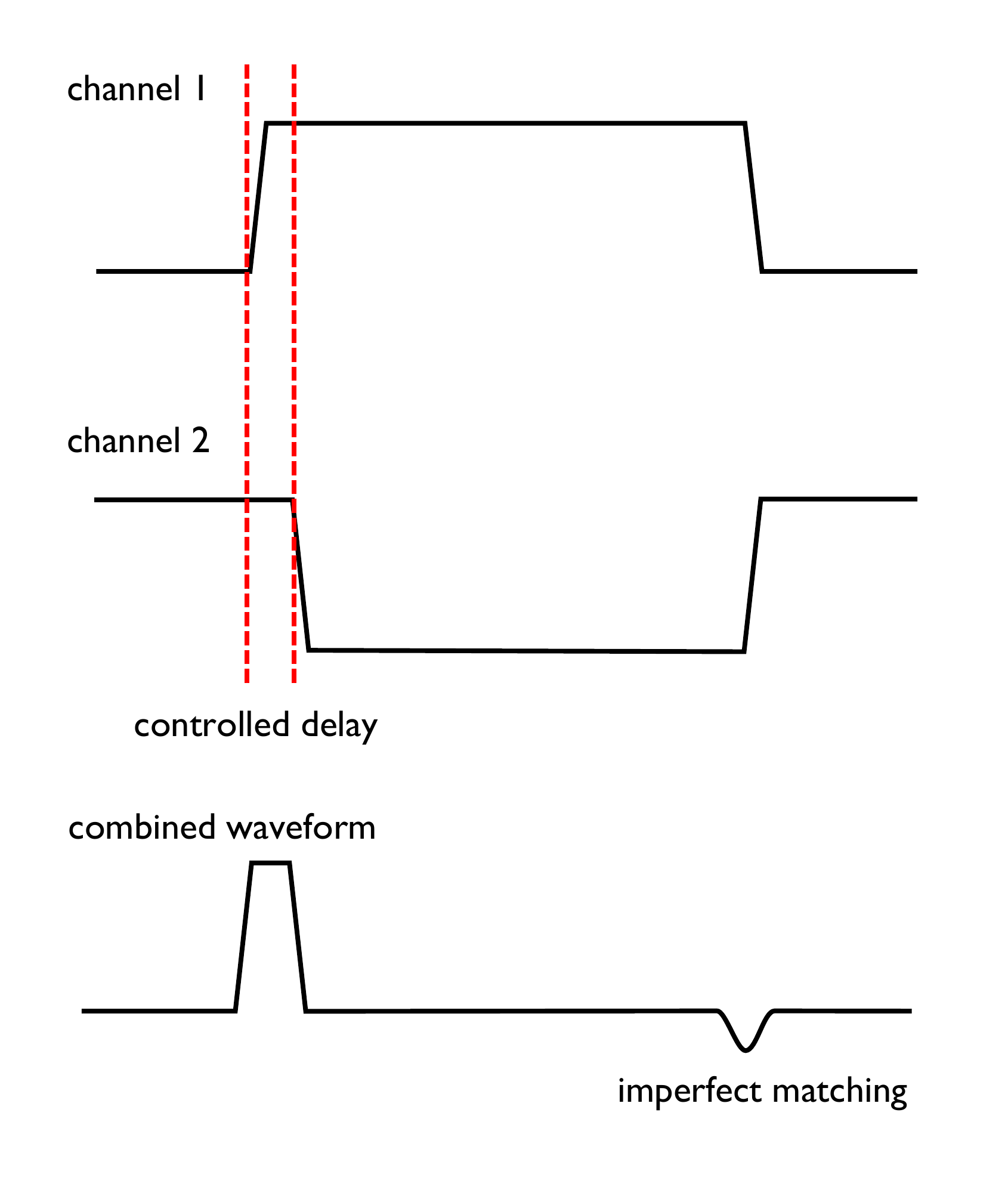}
\caption{\textbf{Pulse generation scheme.} To overcome limitations in the sampling time of the pulse generator, very short pulses were generated by generating a detuning pulse with duration of 10 ns using the first channel of the waveform generator, while a second pulse delayed by $t_\text{p}$with opposite sign and duration 10 ns - $t_\text{p}$ rounded to nearest multiple of 40 ps was generated using the second channel. The pulse width of the combined waveform is controlled by an analog delay of the second channel output, which has timing resolution better than 1 ps. The unwanted detuning variation at the falling edge of the pulse can be minimized modulo the sampling time of the waveform generator (set to 40 ps).}
\label{fig:S2} 
\end{figure}

\begin{figure}[t]
\includegraphics[width=0.47\textwidth]{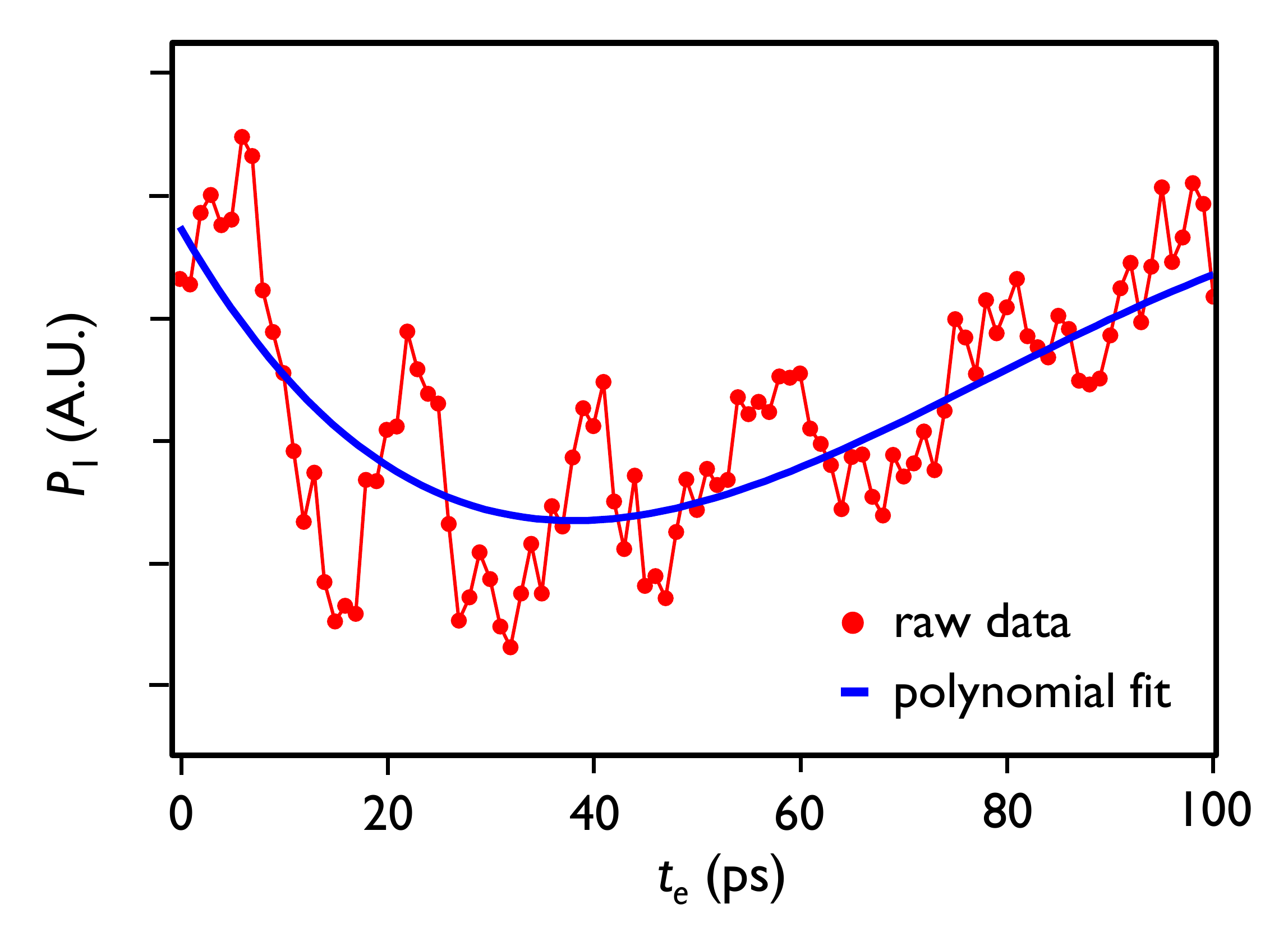}
\caption{\textbf{Background subtraction of Ramsey fringe data.} Coherent $P_{1}$ oscillation as a function of free evolution time $t_{e}$. Blue solid curve shows a fit to a third order polynomial. This smooth
background variation, which is subtracted from the raw data, is likely due to pulse imperfections. }
\label{fig:S3} 
\end{figure}

\begin{figure*}[t]
\includegraphics[width=1\textwidth]{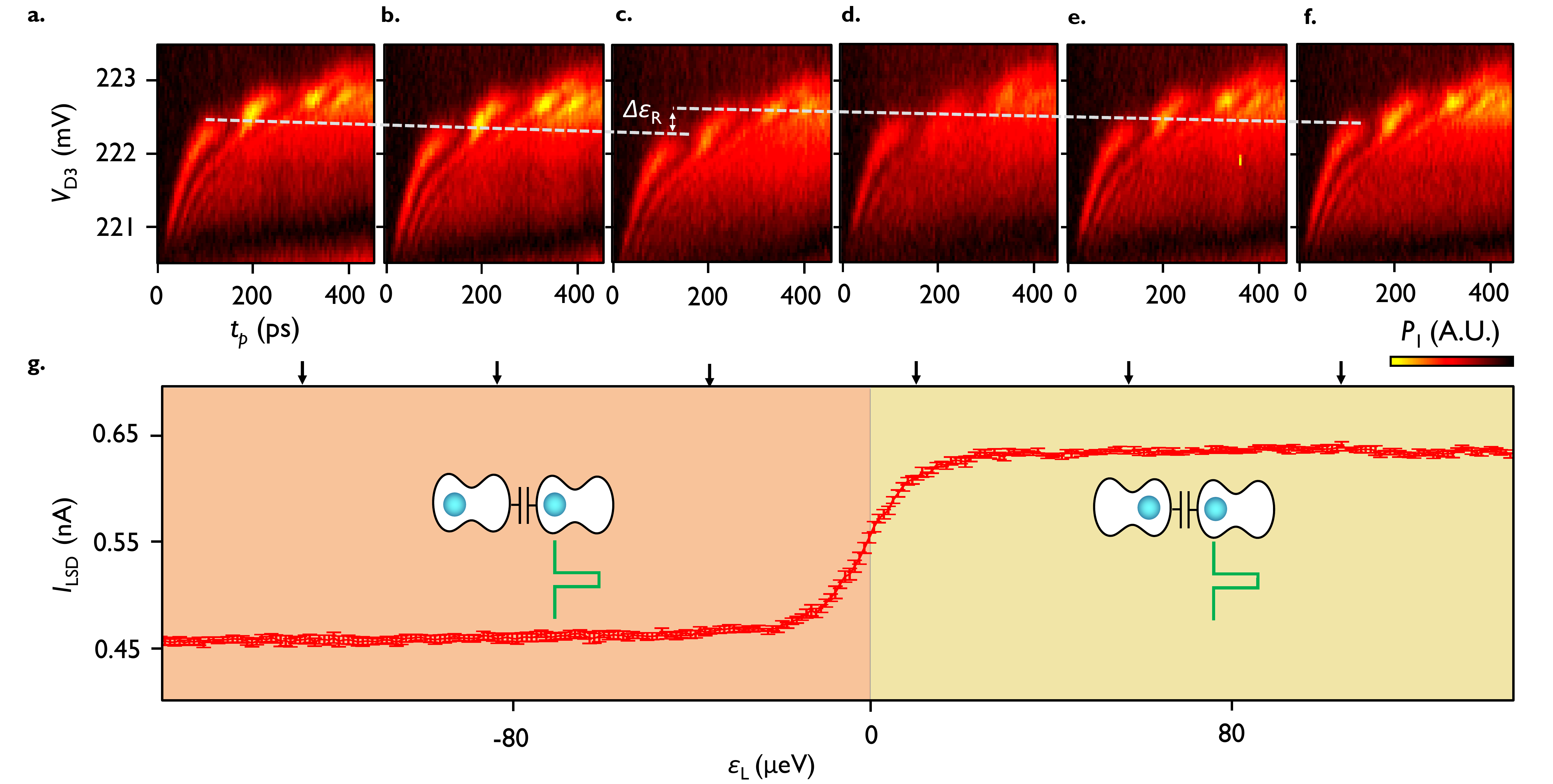}
\caption{\textbf{Variation of right double dot Larmor oscillation patterns as a function of the left double dot detuning. a-f}, $P_\text{1}$ as a function of $V_\text{D3}$ and pulse tip duration $t_\text{p}$ for LDD detuning energy $\varepsilon_\text{L}$ fixed near -100 (\textbf{a}) to +100 $\mu$eV (\textbf{f}). As $\varepsilon_\text{L}$ is changed, the overall Larmor oscillation pattern shifts due to mutual capacitance between gates. However, the Larmor oscillation frequency is roughly constant from \textbf{a} to \textbf{f}. The effect of the LDD charge transition is evident from \textbf{c} to \textbf{d}, where strong capacitive coupling results in a sudden shift of the pattern in the opposite direction (see $\Delta\varepsilon_\text{R}$ in Supplementary Fig.~S4c). \textbf{g}, $I_\text{LSD}$ as a function of LDD energy detuning $\varepsilon_\text{L}$ showing the charge transition near $\varepsilon_\text{L}$=0. Black arrows on top show the values of $\varepsilon_\text{L}$ where the RDD Larmor oscillation experiments in \textbf{a} to \textbf{f} are performed.  }
\label{fig:S4} 
\end{figure*}

\section{Pulse generation and probability measurement details }
\label{sup:pulsegeneration}

Manipulation pulse sequences are generated using a Tektronix AWG70002A arbitrary waveform generator (AWG). The minimum sampling time of 40 ps typically results in timing resolution that is not adequate to measure $>$10 GHz oscillations. To increase timing resolution, we use both channel outputs of the AWG as shown in Supplementary Figure 2. In this approach, channel one outputs a 10 ns duration pulse and channel two outputs an opposite polarity pulse delayed by desired pulse width $t_\text{p}$, which is controlled by an analog skew control with timing resolution better than 1 ps. As the channel output is delayed, a short pulse with opposite polarity appears at about 10 ns due to imperfect cancellation of pulse end edges. Since this opposite pulse is in the measurement step, it does not induce charge transitions. In order to keep the duration of this unwanted mismatch as short as possible, we adjust the total pulse duration of the channel to be 10 ns - $t_\text{p}$ rounded to nearest multiple of 40 ps.  

 In order to measure the charge qubit state probability, we adopt the general scheme described in our previous studies, where we measure the difference between the RSD conductance with and without the manipulation pulses \cite{Kim:2014nature, Kim:2015p243}. The data are acquired using a lock-in amplifier with a reference signal corresponding to the presence and absence of the pulses (lock-in frequency $\approx$ 777\ Hz). We compare the measured signal level with the corresponding $\ket{0}_\text{R}$-$\ket{1}_\text{R}$ charge transition signal level, calibrated by sweeping gate $D_3$ and applying square square pulses with frequency of 777\ Hz to gate $D_3$. 

\section{Background removal in the Ramsey fringe data}
\label{sup:Ramsey}
 Supplementary Fig. S3 shows the raw $P_{1}$ oscillation data corresponding to a Ramsey fringe measurement using two $X_{3\pi/2}$ pulses. The data typically show a smooth background variation in the measured probability, as shown in Supplementary Fig. S3.  This background is likely due to frequency dependent attenuation and inexact calibration of  the state initialization and measurement pulse durations limited by sampling time of the AWG. When we analyze the data to extract the inhomogeneous coherence time $T_{2}$*, we fit the background to third order polynomial (blue solid curve in Supplementary Fig. S3) and subtract this polynomial from the raw data. The resultant high frequency oscillations are shown in Fig. 2g in the main text. 

\section{Charge state dependent Larmor oscillations}
\label{sup:CLarmor}

 In the main text, we showed a coherent interference pattern with fixed RDD energy detuning $\varepsilon_\text{R}$ as the LDD energy detuning $\varepsilon_\text{L}$ is swept. Here we show the interference pattern as a function of $\varepsilon_\text{R}$ with $\varepsilon_\text{L}$ fixed.   Supplementary Figs.~4a-f shows RDD Larmor oscillation patterns as a function of $V_{D3}$ and pulse tip duration $t_p$ as the LDD detuning energy $\varepsilon_\text{L}$ is varied from -100 (\textbf{a}) to +100 $\mu$eV (\textbf{f}). As $\varepsilon_\text{L}$ is changed, the overall coherent oscillation pattern shifts due to mutual capacitance between the gates. However, we find that the Larmor oscillation frequency of about 10 GHz is roughly constant from \textbf{a} to \textbf{f}. The effect of the LDD charge transition is evident when \textbf{c} and \textbf{d} are compared, where the strong capacitive coupling results in a sudden shift of the pattern in the opposite direction (see $\Delta\varepsilon_\text{R}$ in Supplementary Fig.~4c). The origin of this shift is supported by simultaneously measured  $I_\text{LSD}$ (\textbf{g}) as a function of $\varepsilon_\text{L}$, showing that the charge transition indeed occurs in the LDD from \textbf{c} to \textbf{d}, providing evidence that the change in interference pattern frequency measured in the main text is due to a single electron transition.

\bibliography{siliconqcsncwin}

\end{document}